\documentclass [11pt]{article}
\usepackage{color}
\usepackage{epsfig}
\def\beq{\begin{equation}}
\def\eeq{\end{equation}}
\def\beqa{\begin{eqnarray}}
\def\eeqa{\end{eqnarray}}
\def\d{{\rm d}}

\begin{document}
\baselineskip0.55cm plus 1pt minus 1pt
\tolerance=1500

\begin{center}
{\LARGE\bf Forces on wheels and fuel consumption in cars}
\vskip0.4cm
J. G\"u\'emez$^{a,}$\footnote{guemezj@unican.es},
M. Fiolhais$^{b,}$\footnote{tmanuel@teor.fis.uc.pt}
\vskip0.1cm
{\it $^a$ Departamento de F\'{\i}sica Aplicada}\\ {\it Universidad de
Cantabria} \\ {\it E-39005 Santander, Spain}\\
\vskip0.1cm
{\it $^b$ Departamento de F\'\i sica and Centro de
F\'\i sica Computacional}
\\ {\it Universidade de Coimbra}
\\ {\it P-3004-516 Coimbra, Portugal}
\end{center}

\begin{abstract}
Motivated by  real classroom discussions, we analyse the forces acting on moving vehicles, specifically   friction forces on the wheels.
In usual cars, with front-wheel drive, when the car accelerates these forces are in the forward direction in the front wheels, but they
are in the opposite direction in the rear wheels. The situation may be intriguing for students, but it may also be helpful and stimulating to clarify the
role of friction forces on rolling objects. In this article we also study thermodynamical aspects of an accelerating car, relating the distance traveled with the
amount of fuel consumed. The fuel consumption is explicitly shown to be Galilean invariant and we identify the Gibbs free energy as the relevant
quantity that enters in the thermodynamical description of the accelerating car. {\color{black} The more realistic case of the  car's motion
taking into account the dragging forces is also discussed. }
\end{abstract}

\section{Introduction}
\label{sec:intro}

It is not clear when the Mankind firstly used the wheel.  It is even less clear whether the wheel was invented or discovered.
In any case, the wheel has been re-discovered (or re-invented) in several occasions in the course of the times! The re-discovery of the wheel, {\em lato sensu},  also happens in the classroom, sometimes...

Wheels are present everywhere but their most common and evident use is, probably, in vehicles.
Inevitably, when we teach physics, in particular mechanics, sooner or later we are referring to cars, trucks, trains, etc.
Actually,  the reference to common objects always stimulates the attention of the students and helps in passing the idea that what one learns at school is closely related to every day's life.

However, problems may arise when we analyse more deeply some invoked examples. A funny situation may occur when the motion of a car is considered.
 The acceleration of a car is the result of forces exerted by the ground on the tires. Usually, this is recognized by the students but, some of them, may get a bit confused when the teacher classifies  one of those forces as a ``friction force" or, even more adequately, as a ``static friction force".

 --- Isn't it true that the friction force is opposite to the direction of the motion?, one asks.

 --- That is true for  the kinetic friction force acting, for instance, on a point-like object  or in a body in pure translation, but here the wheels are in rotation!, answers the professor. And continues:
 In the case of a wheel, the static friction force can  be either in the forward or in the backward directions with respect to the velocity of the center-of-mass. Another student enters the discussion:

 --- Fine, but, for a car, there shouldn't be any doubt! The  static friction forces on the wheels are, definitely, in the forward direction if the car is increasing its speed, right? Otherwise the car wouldn't accelerate in the forward direction! The professor replies:

 --- Well, that really depends on the price of the car!

 The discussion gets as much amusing as intriguing and the professor has to better explain the point. He/She certainly will think on his/her very old car as a concrete example to help making a clear  statement.

 --- In the case of my car, forward forces are only present on the front wheels  when the car accelerates; on the rear wheels the forces point backwards... And the professor adds:
 This is because I didn't have enough money to buy a car with traction to all four wheels. Such cars are much more expensive, and we, physics teachers, cannot easily afford them!

 In the course of the dialogue the students already get acquainted with a number of interesting points (besides realizing that physics teachers do not own expensive cars): the static friction forces on moving wheels can be either in the direction of the motion or in the opposite direction. Probably they can also vanish (and they do!)... Actually, in cars, one finds friction forces for all tastes!

A rolling cylinder on a horizontal plane, acted upon by a constant horizontal force, $\vec F$, provides a pedagogical example of a system for which the static friction force can be in either direction, depending on the torque produced by $\vec F$. The problem of the identification of the direction of the friction force on rolling objects is well known and ref.~\cite{pinto01} is about this topic.  Many other works touch this or similar problems~\cite{carvalho05},\cite{varios}.

Though some  textbooks address the problem of the forces on the car wheels \cite{halliday01}-\cite{tipler04}, they just do it as a side comment without really solving the problem.
On the other hand, according to our experience as physics teachers, the fact that static friction forces do not produce any work \cite{leff93} is also intriguing, in general, for the students.
Therefore, we also devote some attention to this point.
Actually, since the  static friction forces do not produce any work, the energy required for the car to move does not come from the road... It obviously comes from the fuel and the oxygen in the air. Still using the car as an example, we analyse the dynamical and energy equations, including the first law of thermodynamics. This allows us to formally include in the equations the amount of fuel consumed by the car. We come to the conclusion that it is the free Gibbs energy and not simply the internal energy that matters for the discussion. Moreover, we explore the Galilean Principle of Relativity applied to both the dynamical and energetic equations. Not so surprisingly, the amount of fuel used by the car turns out to be a Galilean invariant. 

The motivation for the present article, {\color{black} whose level of readership is intended for undergraduates}, is, besides the classroom dialogues, the fact that not enough attention is paid to forces acting on the cars, even though cars are constantly used as common examples of moving objects.

In the following section we consider the dynamics of {\color{black} an accelerating} car, representing the forces on the tractive and on the non-tractive wheels. {\color{black} To simplify the discussion we neglect the effects of the air resistance.} In section 3 we address some interesting energy aspects related to a car in motion, {\color{black} using a simplified model to describe the car as a thermodynamical system.} In the same section we also explore the Galilean invariance of the equations describing the processes. {\color{black} In section 4 we introduce the air resistance and relate the fuel consumption with the distance traveled by the car at constant speed. Section 5 is devoted to the conclusions.}

\section{The accelerating car}

A front-wheel drive car of mass $m$ moves with acceleration $a$ (not necessarily constant) 
on a straight horizontal road. The engine produces a
torque $M_0$ which is transmitted to each front wheel by means of a drive shaft (the gearbox function does not need to be considered in our discussion that is about physics and not about mechanical engineering).
Each wheel drive
is acted by a torque  $M_0/2$. The front wheels are also acted upon by other forces that the rest of the car exerts on them, similar and opposite to the forces that the wheels themselves exert on the rest of the car.
These are all internal forces as far as the whole car is concerned. Moreover, the resultant of the vertical forces acting on the car by the ground equals the weight of the car, and both weight and normal forces can be ignored in the present discussion of horizontal motion.  For the sake of clarity we are simplifying the real problem by neglecting air resistance, by assuming that the contact of the wheels with the ground is along a line (and not a surface), etc. In a real classroom context students should be warned about these simplifications. {\color{black} Neglecting the air resistance is a crude approximation, it is only valid when the car starts. However, the simplified situation allows for a clearer discussion. Later on, in section 4, we include the effects of the air resistance.}

\begin{figure}[htb]
\begin{center}
\hspace*{0.5cm}
\includegraphics[width=10cm]{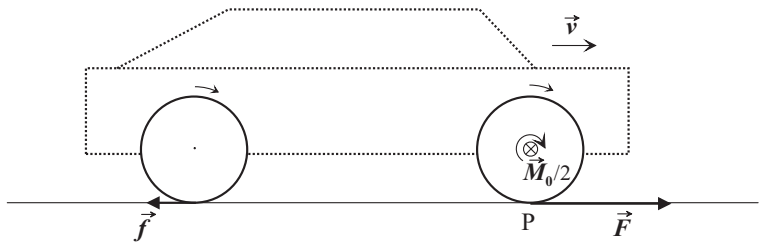}
\end{center}
\vspace*{-0.5cm}
\caption[]{\label{fig:carro} \small Static friction forces acting on the tires.}\vspace*{0.2cm}
\end{figure}

{\color{black} The horizontal forces exerted by the ground on the tires are static friction forces whose magnitude, according to the Amontons-Coulomb law of solid friction cannot exceed $\mu_{\rm s}N$, where
$\mu_{\rm s}$ is the coefficient of static friction and $N$ the vertical normal force exerted by the ground on the tire~\cite{besson07}.  The direction of these static friction forces depends on the type of wheels (tractive or passive) and on whether the acceleration of the car is positive or negative.}
Let us consider one front tractive wheel in {\color{black} accelerated} rotation, as shown in Fig.  \ref{fig:carro}. The contact point with the ground, P, tends to move backwards due to the torque applied to the wheel, so that
the static friction force is in the forward direction. If we denote this friction force by $F$, the rotation equation for the front wheel with respect to the rotation axis is
(the normal force exerted by the ground and the weight of the wheel have zero torque and the rest of the car exerts a force at the center of the wheel, also yielding a vanishing torque)
\beq
{M_0\over 2} -FR= I \alpha\, ,
\label{eq1}
\eeq
where $I$ is the moment of inertia of the wheel, $R$ its radius, and $\alpha=a/R$ (no slipping or smooth rotation condition) is the angular acceleration.

On the other hand, on the rear wheels only the static friction force, denoted by $f$, produces a non-vanishing torque with respect to the center of the wheel. For a wheel to move clockwise, as the rear wheel represented in Fig. \ref{fig:carro},
that force must point backwards! The equation for the rear wheel is, therefore,
\beq
f R= I \alpha\, .
\eeq

Finally, the translation of the car as a whole is governed by the equation
\beq
F_{\rm R}=2F-2f=ma\, ,
\label{eq3}
\eeq
where $F_{\rm R}$ is the resultant  external force.
Altogether, there are three equations, hence three quantities such as $a$, $F$ and $f$ can be determined. From the above equations (\ref{eq1})-(\ref{eq3}), one readily obtains
\beq
\left\{
\begin{array}{rl}
a &=  \frac{R} {4 I + mR^2} \, M_0  \vspace*{0.3cm}\\
F &=  \frac{2I+mR^2} {2 R(4 I + mR^2)} \, M_0  \vspace*{0.3cm}\\
f &=  \frac{I} { R(4 I + mR^2)} \, M_0 \, .
\end{array}
\right.
\label{dfr}
\eeq
 These equations are similar to those given in ref.~\cite{carvalho05}, for the case of a bicycle. {\color{black} In the case of an accelerating bicycle the tractive wheel is the rear one. On that wheel the static
 friction force points forwardly (our $F$ in the car) whereas in the front wheel the static force points in the backward direction (our $f$ in the car). The example or the car and the example of the bicycle, studied in
\cite{carvalho05}, are good and equivalent pedagogical examples to discuss the static friction forces on wheels.}

{\color{black} If we write $I={1\over 2} m' R^2$  (assuming a cylindrical wheel), where $m'$ is the ``effective" wheel mass (actually not the real mass because the wheel is not homogeneous),} the previous equations reduce to
\beq
\left\{
\begin{array}{rl}
a &=  \frac{1} {R (2m' + m)} \, M_0  \vspace*{0.3cm}\\
F &=  \frac{m'+m} {2 R (2m' + m)} \, M_0  \vspace*{0.3cm}\\
f &=  \frac{m'} {2 R (2m'+m) } \, M_0 \, ,
\end{array}
\label{dfr2}
\right.
\eeq
and the resultant force to
\beq
F_{\rm R}=  \frac{m} {(2m' + m)R} \, M_0\, .
\label{rese}
\eeq
\label{sec:boai}

The {\color{black} effective} mass of each wheel is always much smaller than the total mass of the car, $m' \ll m$. From the above equations one concludes that the friction forces on the rear wheels are much smaller than the friction forces on the front wheels, essentially responsible for the acceleration of the car. From equation~(\ref{dfr2})
one obtains, for typical masses of cars and wheels,
\beq
{f \over F} = {m' \over m'+m} \sim 1\%\, .
\eeq
Note that the acceleration and the friction forces are driven by $M_0$ --- see (\ref{dfr2}). In particular, if $M_0=0$, the acceleration and all friction forces also vanish. Of course, we are neglecting the effects of the  air resistance and of the kinetic friction forces always present and, hence, in practice, {\color{black} most} torque communicated by the engine to the tractive wheels
is necessary to keep the car at a constant velocity. {\color{black} This will be discussed in detail in section 4.}

{\color{black} So far, we have considered an accelerating car. In the breaking process, there is a torque applied to the wheels opposite to the torque communicated by the engine. In that case the static friction forces $F$ and $f$ reverse their directions forcing the
centre-of-mass velocity to decrease. }
If the car is four-wheel drive, the torque produced by the engine is distributed by the four wheels and all four static friction forces are in the direction of the velocity of the center-of-mass of the accelerating  car.

\section{Energetic issues}

So far we have analysed forces on the wheels and the acceleration of the car. It is also interesting to
study some energetic aspects  \cite{sousa97} related to the motion of the car. To keep the discussion at the simplest level, let us assume a constant $M_0$. Therefore, the acceleration and the friction forces, given by (\ref{dfr}),  are all constants. From the Newton's second law one obtains the relation between the linear momentum variation of the car and the impulse of the resultant force:
\beq
m v_{\rm cm}= (2F-2f)\, t_0
\label{ret}
\eeq
where $v_{\rm cm}$ is the center-of-mass velocity of the car, initially at rest, after the time $t_0$. Though the static friction forces, $F$ and $f$, do not produce any work, the ``pseudo-work -- kinetic energy" theorem \cite{penchina78} holds and it is expressed, in the present case, by the equation
\beq
{1\over 2} m v_{\rm cm}^2 = (2F-2f) \Delta x_{\rm cm}\, ,
\label{pseudo}
\eeq
where $\Delta x_{\rm cm}$ is the center-of-mass displacement. We stress that the second member of (\ref{pseudo}) is not real work  but rather pseudo-work.

For the rotation of each wheel (front or rear), one has the   following (equivalent) equations \cite{serway02}
\beq
I \omega_{\rm f} = f\, R \, t_0 \, , \ \ \ \ \ \  {1\over 2} I \omega_{\rm f}^2 = fR \theta_0
\label{roteqs}
\eeq
where $\omega_{\rm f}$ is the final angular velocity of the wheels,  $\omega_{\rm f}= v_{\rm cm}/ R$  and $\theta_0=\Delta x_{\rm cm} /R$ (no slipping condition).

Of course, the first law of thermodynamics also applies  and it can  be expressed by the  equation \cite{erlichson84,besson01}
\beq
\Delta {K}_{\rm cm} + \Delta { U} = \sum_j \int { {\vec F}_{{\rm ext}, j}}\cdot \ \d {\vec r}_j   +   Q \, .
\label{totale}
\eeq
This equation is
more general than a center-of-mass equation such as (\ref{pseudo}), but both are valid~\cite{guemez12}.
Each term in the sum on the right-hand side of equation~(\ref{totale})
is work associated with each external force, ${\vec F}_{{\rm ext}, j}$, and  $\d {\vec r}_j$ is the infinitesimal displacement of that  force
(the displacement of the center-of-mass does not enter explicitly in equation~(\ref{totale})). Still on the right-hand side of (\ref{totale}), $Q$ denotes the heat.
On the left-hand side, $\Delta K_{\rm cm}$ denotes the variation of the center-of-mass kinetic energy and $\Delta U$ stands for the sum of {\em all} variations of the internal energy, including rotational energies  or any kinetic energy with respect to the center-of-mass
(besides temperature variation dependencies, work of some internal forces and other possible contributions to the internal energy variations)
\cite{mallin92}.
 In ref.~\cite{sherwood_web} an interesting analysis of a car minus the wheels can be found.

In the present case,  the variation of the internal energy, $\Delta U$, is considered as the sum of just two terms: $\Delta U= \Delta U_{\rm rot}+ \Delta U_\xi$ (we are omitting any internal energy variation from the temperature variation, and all other possible internal energy variations).  The first contribution, $\Delta U_{\rm rot}
= 4 { 1 \over 2} I \omega_{\rm f} ^2$, refers to the
four rotating wheels, all with the same angular velocity. The second contribution is the internal energy variation resulting from the chemical reaction $\xi$ associated with the combustion of the fuel in the engine.
This variation of the internal energy is expressed by
$\Delta U_\xi=n\Delta u_\xi$, where $u_\xi$ is the molar internal energy variation
($\Delta u_\xi$ can, in principle, be obtained from tabulated binding energies, computing the number of broken bonds and the number of formed bonds in the fuel combustion reaction)
and $n$ is the number of moles of fuel burned during the time $t_0$.  Regarding the right-hand side of equation~(\ref{totale}), the work of the external forces is zero for the static friction forces but part of the energy liberated by the chemical reaction $\xi$  is the work, $W_P$, done against the atmospheric pressure, $P$. This work can be expressed as
$W_P=-P\Delta V_\xi=-n P\Delta v_\xi$, where $\Delta V_\xi$ is the associated volume variation. Finally, the heat in the process, in our simplified approach,  can be expressed by $Q=nT \Delta s_\xi$, where $T$ is the temperature of the air (the heat reservoir) and $\Delta s_\xi$ the molar entropy variation for the reaction $\xi$.
Of course, we are oversimplifying the real problem, assuming that the car is just a body with four rotating wheels acted by an engine
where there is a certain chemical reaction.
{\color{black} Our approximation to the thermodynamical description of the car  also does not take into account most of the energy losses to the air and, additionally, we remember that we are not considering the
air resistance in the motion of the car.}
Obviously the car is a much more complicated system but the present model is enough for the pertinent physical discussion {\color{black} of the car in acceleration.}
The bottom line is that equation~(\ref{totale}) leads us to the following energy balance equation  (no  work term related with external forces):
\beq
{1\over 2} m v_{\rm cm}^2 + 4 \, {1\over 2} I \omega_{\rm f}^2 + n \Delta u_\xi = - n P \Delta v_\xi + n T \Delta s_\xi\, .
\label{fgyu}
\eeq
Using the smooth rotation condition and the definition of
the Gibbs's function variation, at constant pressure and temperature, $\Delta g =\Delta u + P \Delta v - T \Delta s$, equation (\ref{fgyu}) can still be written as
\beq
{1 \over 2} \left( m + {4 I \over R^2 }  \right) v_{\rm cm}^2 = - n \Delta g_\xi\, ,
\label{uynm}
\eeq
relating the final velocity of the car with the fuel consumption and with the variation of the   fuel combustion reaction molar Gibbs free energy,  {\color{black} $\Delta g_\xi<0$} (and not the molar internal energy). Using now $v_{\rm cm}= \omega_{\rm f} R$, the second equation (\ref{roteqs}) and the  expression for $f$ in (\ref{dfr}), the equation~(\ref{uynm}) can equivalently be written  as
\beq
M_0\, \theta_0= - n \Delta g_\xi  \, .
\label{m0t0}
\eeq
Using equations (\ref{eq3}) and (\ref{pseudo}) in (\ref{uynm}) the previous two equations may still be written in the form {\color{black}
\beq
\left( 1 + {4 I \over m R^2 }  \right)F_{\rm R} \Delta x_{\rm cm} = -n\Delta g_\xi\, .
\eeq}
%
This equation shows that (the major) part of the Gibbs free energy variation  equals the   energy-like (pseudowork) $F_{\rm R} \Delta x_{\rm cm}$ required to make the car moving forward. In the present model of the car, the other (much smaller) part is used for the rotation of the wheels.

{\color{black} This simplified model of an} accelerating car is a good example of an articulated body with internal sources of energy --   fuel, oxygen  and a thermal engine -- that displaces  its center-of-mass using internal forces ($M_0\theta_0$ in (\ref{m0t0}) is work of internal forces), though the process is intermediated by external forces that do no work, and which result from the contact of the car with an infinite mass body -- the ground -- that remains at rest \cite{mcclelland11}.

\subsection{Principle of Relativity}

All the above equations are valid in the reference frame S of an observer at rest on the road. Both the Newton's second law and the first law of thermodynamics are Galilean invariant, hence there are similar equations for a reference frame S$'$ in standard configuration with respect to S, {\em i.e} moving with velocity   $V$ with respect to S along the common $xx'$ axes. In S$'$, the initial center-of-mass velocity is $-V$ and the final center-of-mass velocity is
$v_{\rm cm}-V$. The center-of-mass displacement  at instant $t_0$ is $\Delta x'_{\rm cm}= \Delta x_{\rm cm}-V t_0$.
Therefore, the equivalent of equations (\ref{ret}) and (\ref{pseudo})  in reference frame S$'$ are
\beq
m (v_{\rm cm}-V)-m  (- V) = (2F-2f)\, t_0
\label{ghfta}
\eeq
and
\beq
{1\over 2} m \left( v_{\rm cm}-V\right)^2 - {1\over 2} mV^2  = (2F-2f) \left(\Delta x_{\rm cm} - V t_0 \right)\, ,
\eeq
respectively. Both equations reduce to the very same equations (\ref{ret}) and (\ref{pseudo}), as expected, according to the Galileo's Principle of Relativity: all inertial reference frames provide exactly the same information --- one cannot get more information in one inertial reference frame with respect to another one. The use of one inertial reference frame or any other one  is usually a matter of a technical advantage. It is important to stress that the rotational equations (\ref{roteqs}) are reference frame invariant.

Regarding equation (\ref{totale}) in S$'$, one should note that the ``static" friction forces now displace their application points by $-Vt_0$: the ground moves with velocity $-V$ in S$'$ and, therefore, they have associated a `force-displacement' product  \cite{hilborn00} (work) given by
$-(2F-2f) \, V\, t_0$.   Hence, equation~(\ref{totale}) in S$'$ leads to
\beq
{1\over 2} m \left( v_{\rm cm}-V\right)^2 - {1\over 2} mV^2+ 4 \, {1\over 2} I \omega_{\rm f}^2=- n' \Delta g_\xi
- (2F-2f) \, V\, t_0
\label{total2}
\eeq
where $n'$ is the number of moles of fuel as measured in S$'$. To write (\ref{total2}) we used the fact that temperature and pressure are Galilean invariants, and all variations of thermodynamical molar quantities, $\Delta u_\xi$, $\Delta v_\xi$ and $\Delta s_\xi$, are also invariants. Moreover, the rotational energy is invariant as well {\color{black} (the rotational equations (\ref{roteqs}) are Galilean invariant)}. Using the result expressed by (\ref{ghfta}) in equation (\ref{total2}), the resulting equation is identical to (\ref{uynm}), as it should, provided $n=n'$. The conclusion is clear: all observers, in any inertial reference frame, agree with the amount of fuel used by the car: $n$ is Galilean invariant. This is an expected result,
however it is  interesting  to get it formally as a consequence of the application of basic Physics laws.

{\color{black}
\section{Air resistance}

The model described so far  provides a good physical description of the car during its acceleration period, if the air resistance is ignored, which is a reasonable approximation only for very low velocities. Hence, in order to obtain a picture closer to reality, we have to take  into account air resistance effects  by including the dragging force applied to the car, $F_{\rm D}$. The dragging force is a function of the density and/or viscosity of the air, of the geometric characteristics of the car, {\em i.e.} of its aerodynamics, and, most important, of the velocity of the car. The detailed form of the dragging force depends crucially on the velocity \cite{benson96}: if the speed is such that there is turbulence behind the car, the dagging force goes as $v_{\rm cm}^2$; for lower velocities  the force depends just on $v_{\rm cm}$ (Stoke's law). We can accommodate these and other regimes in a single expression, by assuming a dragging force of the type
\beq
F_{\rm D}= C\left( {v_{\rm cm} \over v_0} \right)^\alpha\, ,
\label{drag}
\eeq
where the exponent also depends on the centre-of-mass velocity, $\alpha=\alpha(v_{\rm cm})$. In (\ref{drag}) the parameter $v_0$ is  introduced just for dimension purposes and $C$ is a parameter that accounts for the rest of the effects (geometry/aerodynamics, density or viscosity, etc.).

Since the force (\ref{drag}) points backwards, the resultant force on the car is now given by $F_{\rm R}-F_{\rm D}$, where $F_{\rm R}$ is the resultant of the static friction forces introduced in section 2. The equations of motion for the car are now more difficult to derive because of the velocity dependence on $v_{\rm cm}^{\alpha(v_{\rm cm})}$ of the force due to the air resistance. Regarding the energetic issues discussed in section 3, the work associated with the dragging force is a negative dissipative work, $-W_{\rm D}$ (hence, $W_{\rm D}>0$) that contributes to the right hand side of equation (\ref{totale}). The inclusion of this term   changes equation (\ref{uynm}), that now assumes the form
\beq
{1 \over 2} \left( m + {4 I \over R^2 }  \right) v_{\rm cm}^2 + W_{\rm D}= - n \Delta g_\xi\, ,
\label{uyn1}
\eeq
and the conclusion is clear: the fuel consumption in the car or, in more accurate terms, the variation of the Gibbs energy during the acceleration period, is partly used to displace the car (a part of which goes to the rotations of the wheels), and, another part, is ``lost" as dissipative work of the dragging force.

So far we discussed the acceleration of the car. However,  most of the time the car is traveling in a steady  state, at an approximately constant and sufficiently high speed, to make the air resistance a major effect. If the car's velocity is constant, the resultant horizontal force vanishes, $F_{\rm R}-F_{\rm D}=0$, and the forces on the wheels just compensate the dragging force. Let us apply equation (\ref{totale}) in this case. Since the car is in uniform motion, there is no variation of the centre-of-mass kinetic energy, $\Delta K_{\rm cm}=0$, and there is also no internal energy variation associated with the kinetic energy of the wheels, $\Delta U_{\rm rot}=0$. To the internal energy variation associated with the fuel combustion, together with the other aspects related to that chemical reaction, the same discussion presented in section 3 is still applicable here and, altogether, this leads to the term $- n \Delta g_\xi>0$ on the right-hand side of the energy balance equation. On the right-hand side of (\ref{totale}) one has now to include the dissipative work of the dragging force, which is easily computable since the velocity is constant: $-W_{\rm D}=-F_{\rm D}\, \Delta x_{\rm cm}$.
Then, the equation (\ref{totale}) simply reduces to  $F_{\rm D}\, \Delta x_{\rm cm}=- n \Delta g_\xi$ or, using (\ref{drag}) for the dragging force,
\beq
C\left( {v_{\rm cm} \over v_0} \right)^\alpha \Delta x_{\rm cm}= n |\Delta g_\xi|\, .
\label{consumo}
\eeq
This is an interesting equation that relates the fuel consumption with the distance traveled and with the (constant) velocity of the trip. It shows that, in the steady regime, the energy coming from the fuel is all dissipated in the air --- if there were no dragging forces, $C=0$, there were no fuel consumption, $n=0$. It also shows that the fuel consumption, expressed by $n/\Delta x_{\rm cm}$,
\footnote{A quantity equivalent to $n/\Delta x_{\rm cm}$ is commonly referred to, in Europe, as ``liters of gas per hundred kilometers"; in South America a common expression is ``kilometers per liter of gas"; certainly in other regions there are other popular ways to express the fuel consumption in cars.  } strongly depends on the speed of the car. It is worth noting that the time does not enter explicitly in (\ref{consumo}). This means that, if one has to travel a distance $\Delta x_{\rm cm}$, driving slowly is surely the best option to save money. Driving faster always means a larger fuel consumption (also because the exponent $\alpha$ in (\ref{consumo}) increases with the velocity). Even though, shortening the driving time, by increasing the speed, means decreasing the time during which the fuel is consumed, the consumption is unescapably higher if a given distance is traveled at a higher velocity. Therefore, if it happens to the reader getting almost out of gas in the tank, it is advisable not to drive too fast to the next filling station, to arrive earlier, but instead to drive slowly... and eventually arrive!

}

\section{Conclusions}

In conclusion,  a moving car provides examples of friction forces in the direction of the motion and in the opposite direction.
This should not confuse the student. On the contrary, the situation should be explored and regarded as a great opportunity to clarify the role of friction forces in rolling (non-slipping) objects.

An analysis of energetic aspects led us to formally introduce the amount of fuel consumed by the car during the acceleration and relate that amount of fuel with the applied forces and with the distance traveled in every frame. That amount of fuel is reference frame invariant.
{\color{black} Finally, for a car in a steady state motion we related the amount of fuel consumed with the dissipative work of the dragging forces. }

\vspace*{0.5cm}
\noindent {\large\bf Acknowledgement}
\vspace*{0.5cm}

\noindent MF
would like to acknowledge the lively and inspiring discussions with his students of Biomedical Engineering, academic year 2012-2013, during the lectures of ``F\'\i sica~I" (Physics~I).
JG~is grateful to Vidal C. Fern\'andez
for an early discussion on Galilean invariance  in thermodynamics. 


\begin{thebibliography}{99}
\bibitem{pinto01} A. Pinto, M. Fiolhais, {\em Rolling cyllinder on a horizontal plane}, Phys. Edu. {\bf 36} 250-254 (2001)
\bibitem{carvalho05} P. S. Carvalho, A. S. Sousa, {\em Rotation in secondary school:
teaching the effects of frictional force}, Phys. Edu. {\bf 40} 257-265 (2005)
\bibitem{varios} H. Caldas, E. Saltiel,   {\em Les \'etudiants et les forces de frottement solide: le mod\`ele de la
brosse}, Le BUP {\bf 822} 471–485 (2000); U. Besson, L. Borghi, A. Ambrosis, P. Mascheretti, {\em A Three-Dimensional Approach and Open Source Structure for the Design and Experimentation of Teaching-Learning Sequences: The case of friction}, Int. J. Sci. Educ. {\bf 32}
1289-1313 (2010).
\bibitem{halliday01} D. Halliday, R. Resnick, J. Walker, {\it Fundamentals of Physics}, 6th Ed. John Wiley and Sons, New York (2001)
\bibitem{walker01} J. Walker, D.   Halliday, R. Resnick, {\it Principles of Physics. International Student Version}, 9th Ed. John Wiley and Sons, New York (2001)
\bibitem{benson96} H. Benson, {\it University Physics}, Revised edition, John Wiley and Sons, New York (1996)
\bibitem{tipler04} P. A. Tipler, G. Mosca, {\it Physics for Scientists and Engineers}, 5th Ed. W H Freeman and Co., New York (2004)
\bibitem{leff93}  H. S. Leff, A.  J.  Mallinckrodt, {\it Stopping objects with zero external work: Mechanics meets thermodynamics}, Am.  J.  Phys.   {\bf 61}   121-127 (1993)
{\color{black}
\bibitem{besson07} U. Besson, L. Borghi, A. De Ambrosis, P. Mascheretti, {\it How to teach friction: Experiments and models}, Am. J. Phys. {\bf 75} 1106-1113 (2007)
}
\bibitem{sousa97}  C. A. Sousa, E. P. Pina, {\em Aspects of mechanics and thermodynamics in introductory physics: an illustration in the context of fiction and rolling}, Eur. J. Phys. {\bf 19} 334-337 (1997).
\bibitem{penchina78} C.  M.  Penchina, {\it Pseudowork-energy principle},    Am. J. Phys. {\bf 46} 295-296 (1978)
\bibitem{serway02} R. A. Serway, J. W. Jewett, Jr., {\it Principles of Physics. A Calculus-Based Text},    3rd Ed. Harcourt College Publishers, Fort Worth (2002)
\bibitem{erlichson84}  H. Erlichson, {\it Internal energy in the first law of thermodynamics}, Am. J. Phys. {\bf 52}  623-625 (1984)
\bibitem{besson01} U.  Besson, {\it Work and energy in the presence of friction: the need for a mesoscopic analysis}, Eur. J. Phys. {\bf 22}  613-622 (2001)
\bibitem{guemez12} J. G\"u\'emez, M. Fiolhais, {\em From mechanics to thermodynamics: analysis of selected examples}, Eur. J. Phys.  {\bf 34} 345-357 (2013)
{\color{black}
\bibitem{mallin92}  A.   J.  Mallinckrodt, H.  S.  Leff, {\it All about work},  Am.  J.  Phys.   {\bf 60}    356-365 (1992);
A. J. Mallinckrodt, H. S. Leff, {\em Work, Energy, and Kinematic Car Data}, Phys. Teach. {\bf 40} 516-517 (2002)
\bibitem{sherwood_web} B. Sherwood, {\em Work and energy for an accelerating car}, \\
{\small http://matterandinteractions.wordpress.com/2012/01/09/work-and-energy-for-an-accelerating-car/}
}
\bibitem{mcclelland11} J. A. G. McClelland, {\it A very persistent mistake}, Phys. Educ. {\bf 46} 469-471 (2011)
\bibitem{hilborn00}  R. C. Hilborn, {\it Let's ban Work from Physics},    Phys. Teach. {\bf 38} 447 (2000)

\end{thebibliography}
\end{document}